\documentclass[%
 aip,
 jmp,%
 amsmath,amssymb,
 reprint,%
]{revtex4-1}

\usepackage{braket}
\usepackage{mathtools}
\usepackage{graphicx}% Include figure files
\usepackage{dcolumn}% Align table columns on decimal point
\usepackage{bm}% bold math
\usepackage[breaklinks=true,colorlinks=true,linkcolor=blue,urlcolor=blue,citecolor=blue]{hyperref}

\usepackage{xcolor}

\newcommand{\op}[1]{\ensuremath{\hat{#1}}}
\newcommand{\ladderdown}{\ensuremath{\op{a}^{\vphantom{\dagger}}}}
\newcommand{\ladderup}{\ensuremath{\op{a}^\dagger}}
\newcommand{\annihilop}{\ladderdown}
\newcommand{\creationop}{\ladderup}

\begin{document}

\preprint{AIP/123-QED}

\title[]{\emph{Ab initio} Quantum Monte Carlo simulation of the warm dense electron gas}% Force line breaks with \\
%\thanks{Footnote to title of article.}

\author{Tobias Dornheim} 
 \email{dornheim@theo-physik.uni-kiel.de}
 \affiliation{Institut f\"ur Theoretische Physik und Astrophysik, Christian-Albrechts-Universit\"{a}t zu Kiel, D-24098 Kiel, Germany}

\author{Simon Groth} \altaffiliation[ ]{TD and SG contributed equally to this work.}
\affiliation{Institut f\"ur Theoretische Physik und Astrophysik, Christian-Albrechts-Universit\"{a}t zu Kiel, D-24098 Kiel, Germany}
\author{Fionn D.~Malone}\affiliation{Department of Physics, Imperial College London, Exhibition Road, London SW7 2AZ, United Kingdom}
\author{Tim Schoof}%
\affiliation{Institut f\"ur Theoretische Physik und Astrophysik, Christian-Albrechts-Universit\"{a}t zu Kiel, D-24098 Kiel, Germany}
\author{Travis Sjostrom}\affiliation{Theoretical Division, Los Alamos National Laboratory, Los Alamos, New Mexico 87545, USA}
\author{W.M.C.~Foulkes}\affiliation{Department of Physics, Imperial College London, Exhibition Road, London SW7 2AZ, United Kingdom}
\author{Michael Bonitz}
\affiliation{Institut f\"ur Theoretische Physik und Astrophysik, Christian-Albrechts-Universit\"{a}t zu Kiel, D-24098 Kiel, Germany}

\date{\today}% It is always \today, today,

\begin{abstract}
Warm dense matter is one of the most active frontiers in plasma physics due to its relevance for dense astrophysical objects as well as for novel laboratory experiments in which matter is being strongly compressed e.g. by high-power lasers. Its description is theoretically very challenging as it contains correlated quantum electrons at finite temperature---a system that cannot be accurately modeled by standard analytical or ground state approaches. Recently several breakthroughs have been achieved in the field of fermionic quantum Monte Carlo simulations. First, it was shown that exact simulations of a finite model system ($30 \dots 100$ electrons) is possible that avoid any simplifying approximations such as fixed nodes [Schoof {\em et al.}, Phys. Rev. Lett. {\bf 115}, 130402 (2015)]. Second, a novel way to accurately extrapolate these results to the thermodynamic limit was reported by Dornheim {\em et al.} [Phys. Rev. Lett. {\bf 117}, 156403 (2016)]. As a result, now thermodynamic results for the warm dense electron gas are available that have an unprecedented accuracy on the order of $0.1\%$.  
Here we present an overview on these results and discuss limitations and future directions.
%
%Valid PACS numbers may be entered using the \verb+\pacs{#1}+ command.
\end{abstract}

\pacs{Valid PACS appear here}% PACS, the Physics and Astronomy
%Use showkeys class option if keyword
\maketitle

\section{\label{sec:introduction}Introduction}

The uniform electron gas (UEG) (i.e., electrons in a uniform positive background) is inarguably one of the most fundamental systems in condensed matter physics and quantum chemistry\cite{loos}. Most notably, the availability of accurate quantum Monte Carlo (QMC) data for its ground state properties\cite{alder,perdew} has been pivotal for the success of Kohn-Sham density functional theory (DFT) \cite{ks,dft_review}.

Over the past few years, interest in the study of matter under extreme conditions has grown rapidly.
Experiments with inertial confinement fusion targets \cite{nora,schmit,hurricane3} and laser-excited solids \cite{ernst}, but also astrophysical applications such as planet cores and white dwarf atmospheres \cite{knudson,militzer}, require a fundamental understanding of the warm dense matter (WDM) regime, a problem now at the forefront of plasma physics and materials science. 
In this peculiar state of matter, both the dimensionless Wigner-Seitz radius $r_s=\overline{r}/a_0$ (with the mean interparticle distance $\overline{r}$ and Bohr radius $a_0$) and the reduced temperature $\theta=k_\textnormal{B} T/E_\textnormal{F}$ ($E_\textnormal{F}$ being the Fermi energy) are of order unity, implying a complicated interplay of quantum degeneracy, coupling effects, and thermal excitations. Because of the importance of thermal excitation, the usual ground-state version of DFT does not provide an appropriate description of WDM. 
An explicitly thermodynamic generalization of DFT\cite{mermin} has long been known in principle, but requires an accurate parametrization of the exchange-correlation free energy ($f_{xc}$) of the UEG over the entire warm dense regime as an input\cite{karasiev2,dharma,gga,burke,burke2}.

This seemingly manageable task turns out to be a major obstacle. The absence of a small parameter prevents a low-temperature or perturbation expansion and, consequently, Green function techniques in the Montroll-Ward and $e^4$ approximations\cite{kremp_springer,vorberger_pre04} break down.  
Further, linear response theory within the random phase approximation\cite{gupta,pdw84} (RPA) and also with the additional inclusion of static local field corrections as suggested by, e.g., Singwi, Tosi, Land, and Sj\"olander\cite{stls_original,stls,stls2} (STLS) and Vashista and Singwi\cite{vs_original,stls2} (VS), are not reliable. Quantum classical mappings\cite{rpa0,dwp} are exact in some known limiting cases, but constitute an uncontrolled approximation in the WDM regime.

The difficulty of constructing a quantitatively accurate theory of WDM leaves thermodynamic QMC simulations as the only available tool to accomplish the task at hand. However, the widely used path integral Monte Carlo\cite{cep} (PIMC) approach is severely hampered by the notorious fermion sign problem\cite{loh,troyer} (FSP), which limits simulations to high temperatures and low densities and precludes applications to the warm dense regime. In their pioneering work, Brown \textit{et al.}\cite{brown} circumvented the FSP by using the fixed-node approximation\cite{node} (an approach hereafter referred to as restricted PIMC, RPIMC), which allowed them to present the first comprehensive results for the UEG over a wide temperature range for $r_s\geq1$.

Although these data have subsequently been used to construct the parametrization of $f_{xc}$ required for thermodynamic DFT (see Refs.~\onlinecite{brown3,karasiev,stls2}), their quality has been called into question.
Firstly, RPIMC constitutes an uncontrolled approximation\cite{vfil1,vfil2,hyd1,hyd2}, which means that the accuracy of the results for the finite model system studied by Brown \textit{et al.}\cite{brown} was unclear. This unsatisfactory situation has sparked remarkable recent progress in the field of fermionic QMC\cite{vfil4,blunt,malone,malone2,dornheim,dornheim2,dornheim_prl,dornheim_cpp,tim_prl,tim1,tim2,blunt2}. In particular, a combination of two complementary QMC approaches\cite{dornheim3,groth} has recently been used to simulate the warm dense UEG without nodal restrictions\cite{tim_prl}, revealing that the nodal contraints in RPIMC cause errors exceeding $10\%$. 
Secondly, Brown \textit{et al.}\cite{brown} extrapolated their QMC results for $N=33$ spin-polarized ($N=66$ unpolarized) electrons to the macroscopic limit by applying a finite-$T$ generalization of the simple first-order finite-size correction (FSC) introduced by Chiesa \textit{et al.}\cite{chiesa} for the ground state. As we have recently shown\cite{dornheim_prl}, this is only appropriate for low temperature and strong coupling and, thus, introduces a second source of systematic error.

In this paper, we give a concise overview of the current state of the art of quantum Monte Carlo simulations of the warm dense electron gas and present new results regarding the extrapolation to the thermodynamic limit. Further, we discuss the remaining open questions and challenges in the field. 

 After a brief introduction to the UEG model (\ref{Sec:ms}) we introduce various QMC techniques, starting with the standard path integral Monte Carlo approach (\ref{sub:pimc}) and a discussion of the origin of the FSP (\ref{sub:sign_problem}).
The sign problem can be alleviated using either the permutation blocking PIMC (PB-PIMC, \ref{sub:pbpimc}) method, or the configuration PIMC algorithm (CPIMC, \ref{sub:cpimc}), or the density matrix QMC (DMQMC, \ref{sub:dmqmc}) approach.
In combination, these tools make it possible to obtain accurate results for a finite model system over almost the entire warm dense regime (\ref{sec:finite}).
The second key issue is the extrapolation from the finite to the infinite system, i.e., the development of an appropriate finite-size correction\cite{chiesa,lin,drummond,fraser,krakauer,dornheim_prl}, which is discussed in detail in Sec.~\ref{sec:finite_size}.          
Finally, we compare our QMC results for the infinite UEG to other data (\ref{ss_compare}) and finish with some concluding remarks and a summary of open questions.

\section{The Uniform Electron Gas\label{Sec:ms}}
\subsection{Coordinate representation of the Hamiltonian\label{cor}}
Following Refs.~\onlinecite{fraser,dornheim2}, we express the Hamiltonian (using Hartree atomic units) for $N=N_\uparrow+N_\downarrow$ unpolarized electrons in coordinate space as 
 \begin{eqnarray}
  \hat{H} = - \frac{1}{2}\sum_{i=1}^N \nabla^2_i +  \frac{1}{2}\sum_{i=1}^N\sum_{j\ne i}^N \Psi( \mathbf{r}_i, \mathbf{r}_j) + \frac{ {N} }{2}\xi_\textnormal{M} \; ,
  \label{Hcoord}
 \end{eqnarray}
with the well-known Madelung constant $\xi_\textnormal{M}$ and the periodic Ewald pair interaction
\begin{eqnarray}
\Psi(\mathbf{r}, \mathbf{s} ) &=& \frac{1}{\Omega} \sum_{ \mathbf{G} \ne 0 } \frac{ e^{-\pi^2\mathbf{G}^2/\kappa^2} e^{2\pi i \mathbf{G}(\mathbf{r}-\mathbf{s})} }{ \pi\mathbf{G}^2}
  \label{pair} \nonumber \\ &-& \frac{\pi}{\kappa^2 \Omega} + \sum_\mathbf{R} \frac{ \textnormal{erfc}( \kappa | \mathbf{r}-\mathbf{s} + \mathbf{R} | ) }{ |\mathbf{r}-\mathbf{s}+\mathbf{R} | } \ .
\end{eqnarray}
Here $\mathbf{R}=\mathbf{n}_1L$ and $\mathbf{G}=\mathbf{n}_2/L$ denote the real and reciprocal space lattice vectors, respectively, with $\mathbf{n}_1$ and $\mathbf{n}_2$ three-component vectors of integers, $L$ the box length, $\Omega=L^3$ the box volume, and $\kappa$ the usual Ewald parameter.

 \subsection{Hamiltonian in second quantization\label{sec:second_quant}}
In second quantized notation using a basis of spin-orbitals of plane waves,
$\langle \mathbf{r} \sigma \;|\mathbf{k}_i\sigma_i\rangle = \frac{1}{L^{3/2}} e^{i\mathbf{k}_i \cdot \mathbf{r}}\delta_{\sigma,\sigma_i}$, with $\mathbf{k}_i=\frac{2\pi}{L}\mathbf{m}_i$, $\mathbf{m}_i\in \mathbb{Z}^3$ and $\sigma_i\in\{\uparrow,\downarrow\}$, 
the Hamiltonian, Eq.~(\ref{Hcoord}), becomes
\begin{align}\label{eq:h} 
& \op{H} =
%\begin{aligned}[t]
\frac{1}{2}\sum_{i}\mathbf{k}_i^2 \creationop_{i}\annihilop_{i} +
\smashoperator{\sum_{\substack{i<j,k<l \\ i\neq k,j\neq l}}} 
w^-_{ijkl}\creationop_{i}\creationop_{j} \annihilop_{l} \annihilop_{k} + \frac{N}{2}\xi_M.
\end{align}
% \textcolor{red}{[WMCF: please check change to Madelung term here.]}
The antisymmetrized two-electron integrals take the form $w^-_{ijkl} =w_{ijkl}-w_{ijlk}$, where
\begin{align} 
\; w_{ijkl}=\frac{4\pi e^2}{L^3 (\mathbf{k}_{i} - \mathbf{k}_{k})^2}\delta_{\mathbf{k}_i+\mathbf{k}_j, \mathbf{k}_k + \mathbf{k}_l}\delta_{\sigma_i,\sigma_k}\delta_{\sigma_j,\sigma_l}\ ,
\label{eq:two_ints}
\end{align}
and the Kronecker deltas ensure both momentum and spin conservation. The first (second) term in the Hamiltonian, Eq.~(\ref{eq:h}), describes the kinetic (interaction) energy.
%(In this paper Rydberg units are used.). 
The operator 
$\creationop_{i}$  ($ \annihilop_{i}$) 
%$\creationop_{\mathbf{k}_i}$  ($ \annihilop_{\mathbf{k}_i}$) 
creates (annihilates) a particle in the spin-orbital $|\mathbf{k}_i\sigma_i\rangle$.

\section{\label{sec:qmc}Quantum Monte Carlo}
\subsection{\label{sub:pimc}Path integral Monte Carlo}
Let us consider $N$ spinless distinguishable particles in the canonical ensemble, with the volume $\Omega$, the inverse temperature $\beta=1/k_\textnormal{B}T$, and the density $N/\Omega$ being fixed. The partition function in coordinate representation is given by 
\begin{eqnarray}
 Z =  \int \textnormal{d}\mathbf{R}\ \bra{ \mathbf{R} } e^{-\beta\hat{H}} \ket{ \mathbf{R}} \quad , \label{Z}
\end{eqnarray}
where $\mathbf{R}=\{\mathbf{r}_1,\dots,\mathbf{r}_N\}$ contains all $3N$ particle coordinates, and the Hamiltonian $\hat{H}=\hat{K}+\hat{V}$ is given by the sum of a kinetic and a potential part, respectively. Since the low-temperature matrix elements of the density operator, $\hat{\rho}=e^{-\beta\hat{H}}$, are not readily known, we exploit the group property $e^{-\beta\hat{H}}=\left( e^{-\epsilon\hat{H}} \right)^P$, with $\epsilon=\beta/P$ and positive integers $P$. Inserting $P-1$ unities of the form $\hat{1} = \int \textnormal{d}\mathbf{R}_\alpha\ \ket{\mathbf{R}}_\alpha \bra{\mathbf{R}}_\alpha$ into Eq.~(\ref{Z}) leads to
\begin{eqnarray} \nonumber
 Z &=& \int  \textnormal{d}\mathbf{X}\ \left( \bra{\mathbf{R}_0} e^{-\epsilon\hat{H}} \ket{\mathbf{R}_1} \bra{\mathbf{R}_1} \dots \ket{\mathbf{R}_{P-1}} \right. \\  & &\left. \bra{\mathbf{R}_{P-1}} e^{-\epsilon\hat{H}} \ket{\mathbf{R}_0} \right)  \label{group} 
 = \int  \textnormal{d}\mathbf{X}\ W(\mathbf{X}) \quad .
\end{eqnarray}
We stress that Eq.~(\ref{group}) is still exact and constitutes an integral over $P$ sets of particle coordinates ($\textnormal{d}\mathbf{X}=\textnormal{d}\mathbf{R}_0\dots\textnormal{d}\mathbf{R}_{P-1}$), the integrand being a product of $P$ density matrices, each at $P$ times the original temperature $T$.
Despite the significantly increased dimensionality of the integral, this recasting is advantageous as the high temperature matrix elements can easily be approximated, most simply with the primitive approximation, $e^{-\epsilon\hat{H}}\approx e^{-\epsilon\hat{K}}e^{-\epsilon\hat{V}}$, which becomes exact for $P\to\infty$.
In a nutshell, the basic idea of the path integral Monte Carlo method \cite{cep} is to map the quantum system onto a classical ensemble of ring polymers \cite{chandler}. The resulting high dimensional integral is evaluated using the Metropolis algorithm \cite{metropolis}, which allows one to sample the $3PN$-dimensional configurations $\mathbf{X}$ of the ring polymer according to the corresponding configuration weight $W(\mathbf{X})$.

\subsection{\label{sub:sign_problem}The fermion sign problem}
To simulate $N$ spin-polarized fermions, the partition function from the previous section has to be extended to include a sum over all $N!$ permutations of particles:
\begin{eqnarray}
 Z = \frac{1}{N!} \sum_{s\in S_N} \textnormal{sgn}(s) \int \textnormal{d}\mathbf{R}\ \bra{ \mathbf{R} } e^{-\beta\hat{H}} \ket{ \hat{\pi}_s\mathbf{R}} \quad , \label{fermi_Z}
\end{eqnarray}
% \textcolor{red}{[WMCF: earlier on, $\sigma$ was the spin, so your notation for permutations might be confusing.]}
    where $\hat{\pi}_s$ denotes the exchange operator corresponding to the element $s$ from the permutation group $S_N$. Evidently, Eq.~(\ref{fermi_Z}) constitutes a sum over both positive and negative terms, so tht the configuration weight function $W(\mathbf{X})$ can no longer be interpreted as a probability distribution. To allow fermionic expectation values to be computed using the Metropolis Monte Carlo method, we introduce the modified partition function 
\begin{eqnarray}
Z' = \int  \textnormal{d}\mathbf{X}\ |W(\mathbf{X})| \quad ,
\end{eqnarray}
and compute fermionic observables as
\begin{align}
\langle O \rangle = \frac{\langle OS\rangle^\prime}{\langle S \rangle^\prime}\;, 
\label{eq:average}
\end{align}
with averages taken over the modified probability distribution $W'(\mathbf{X}) = |W(\mathbf{X})|$ and $S=W(\mathbf{X})/|W(\mathbf{X})|$ denoting the sign.
The average sign, i.e., the denominator in Eq.~(\ref{eq:average}),
is a measure for the cancellation of positive and negative contributions and exponentially decreases with inverse temperature and system size, $\braket{S}'\propto e^{-\beta N(f-f')}$, with $f$ and $f'$ being the free energy per particle of the original and the modified system, respectively.
The statistical error of the Monte Carlo average value $\Delta O$ is inversely proportional to $\braket{S}'$,
\begin{align}
\frac{\Delta O}{O} \propto \frac{1}{\braket{S}' \sqrt{N_\textnormal{MC}} } \propto \frac{ e^{\beta N (f-f')}}{ \sqrt{N_\textnormal{MC}}} \label{eq:FSP} \quad .
\end{align}
The exponential increase of the statistical error with $\beta$ and $N$ evident in Eq.~(\ref{eq:FSP}) can only be compensated by increasing the number of Monte Carlo samples, but the slow $1/\sqrt{N_\textnormal{MC}}$ convergence soon makes this approach unfeasible. 
This is the notorious fermion sign problem \cite{loh,troyer}, which renders standard PIMC unfeasible even for the simulation of small systems at moderate temperature.

\subsection{\label{sub:pbpimc}Permutation blocking path integral Monte Carlo}
To alleviate the difficulties associated with the FSP, Dornheim \textit{et al.}\cite{dornheim,dornheim2,dornheim_cpp} recently introduced a novel simulation scheme that significantly extends fermionic PIMC simulations towards lower temperature and higher density. This so-called permutation blocking PIMC (PB-PIMC) approach combines: (i) the use of antisymmetrized density matrix elements, i.e., determinants \cite{det1,det2,det3}; (ii) a fourth-order factorization scheme to obtain accurate approximate density matrices for relatively low temperatures (large imaginary-time steps) \cite{ho1,ho2,ho3,ho4}; and (iii) an efficient Metropolis Monte Carlo sampling scheme based on the temporary construction of artificial trajectories \cite{dornheim}.

In particular, we use the factorization introduced in Refs.~\onlinecite{ho3,ho2}
\begin{eqnarray}
 \label{eq:chin} e^{-\epsilon\hat{H}} \approx e^{-v_1\epsilon\hat{W}_{a_1}} e^{-t_1\epsilon\hat{K}} e^{-v_2\epsilon\hat{W}_{1-2a_1}} \\ \nonumber e^{-t_1\epsilon\hat{K}} e^{-v_1\epsilon\hat{W}_{a_1}} e^{-2t_0\epsilon\hat{K}} \quad ,
\end{eqnarray}
where the $\hat{W}$ operators denote a modified potential term, which combines the usual potential energy $\hat{V}$ with
double commutator terms of the form 
\begin{eqnarray}
 [[\hat{V},\hat{K}],\hat{V}] = \frac{\hbar^2}{m} \sum_{i=1}^N |\mathbf{F}_i|^2 \quad ,
\end{eqnarray}
and, thus, requires the evaluation of all forces in the system. Furthermore, for each high-temperature factor, there appear three imaginary time steps.
The final result for the partition function is given by 
\begin{eqnarray}
\label{eq:final_Z} Z = \frac{1}{(N!)^{3P}} \int \textnormal{d}\mathbf{X} \prod_{\alpha=0}^{P-1} \bigg( e^{-\epsilon\tilde V_\alpha}e^{-\epsilon^3u_0\frac{\hbar^2}{m}\tilde F_\alpha}  \\ \nonumber  \textnormal{det}(\rho_\alpha)\textnormal{det}(\rho_{\alpha A})\textnormal{det}(\rho_{\alpha B}) \bigg) \quad ,
\end{eqnarray}
where the determinants incorporate the three diffusion matrices for each of the $P$ factors\cite{dornheim2},
\begin{eqnarray}
 \rho_\alpha(i,j) &=& \lambda_{t_1\epsilon}^{-D} \sum_{\mathbf{n}} \textnormal{exp}{\left( -\frac{\pi ( \mathbf{r}_{\alpha,j} - \mathbf{r}_{\alpha A,i} + \mathbf{n}L)^2}{\lambda^2_{t_1\epsilon}} \right)} \ . \label{eq:diffusion}
\end{eqnarray}

The key problem of fermionic PIMC simulations is the sum over permutations, where each configuration can have a positive or a negative sign. By introducing determinants, we analytically combine both positive and negative contributions into a single configuration weight (hence the label 'permutation blocking'). Therefore, parts of the cancellation are carried out beforehand and the average sign of our simulations [Eq.~(\ref{eq:average})] is significantly increased. Since this effect diminishes with increasing $P$, we employ the fourth-order factorization, Eq.~(\ref{eq:chin}), to obtain sufficient (although limited~\cite{dornheim2}, $|\Delta V| / V\lesssim 0.1\%$) accuracy with only a small number of high-temperature factors. PB-PIMC is a substantial improvement over regular PIMC, but the determinants can still be negative, which means that the FSP is not removed by the PB-PIMC approach. 
\begin{figure}[]
 \centering
\includegraphics[width=0.44\textwidth]{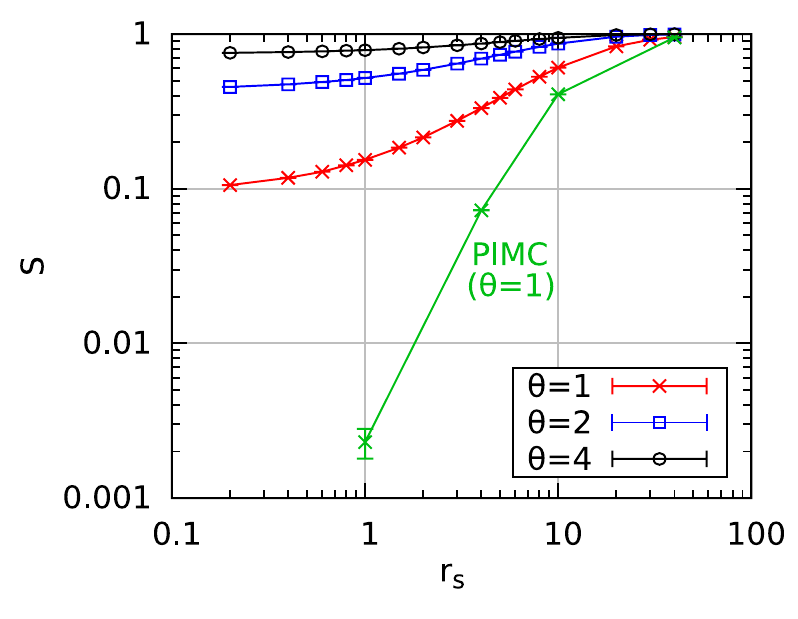}\vspace*{-0.25cm}
 \caption{\label{fig:jcp1} 
Density dependence of the average sign of a PB-PIMC simulation of the uniform electron gas. Also shown are standard PIMC data taken from Ref.~\onlinecite{brown}. The figure has been taken from Dornheim \textit{et al.}~\cite{dornheim2}.
}
\end{figure}
To illustrate this point, in Fig.~\ref{fig:jcp1} we show simulation results for the average sign (here denoted as $S$) as a function of the density parameter $r_s$ for a UEG simulation cell containing $N=33$ spin-polarized
electrons subject to periodic boundary conditions. The red, blue, and black curves correspond to PB-PIMC results for three isotherms and exhibit a qualitatively similar behavior. At high $r_s$, fermionic exchange is suppressed by the strong Coulomb repulsion, which means that almost all configuration weights are positive and $S$ is large. With increasing density, the system becomes more ideal and the electron wave functions overlap, an effect that manifests itself in an increased number of negative determinants. Nevertheless, the value of $S$ remains significantly larger than zero, which means that, for the three depicted temperatures, PB-PIMC simulations are possible over the entire density range.
In contrast, the green curve shows the density-dependent average sign for standard PIMC simulations\cite{brown} at $\theta=1$ and exhibits a significantly steeper decrease with density, limiting simulations to $r_s\geq4$.

%\begin{figure}[]
% \centering
%\includegraphics[width=0.41\textwidth]{dornheim_njp1.png}\vspace*{-0.25cm}
% \caption{\label{fig:njp1} 
%Schematical illustration of the path integral representation and the idea of permutation blocking. The %figure has been taken from Dornheim \textit{et al.}~\cite{dornheim}.
%}
%\end{figure}

\subsection{\label{sub:cpimc}Configuration path integral Monte Carlo}
%\begin{figure}
%\includegraphics[width=85mm]{cpimc_sketch_unpol.pdf}
% \caption{Typical closed path of $N=4$ unpolarized particles in Slater determinant (Fock) space. The state with four occupied orbitals $|\mathbf{k}_0\uparrow\rangle, |\mathbf {k}_1\downarrow\rangle, |\mathbf {k}_3\downarrow\rangle, |\mathbf {k}_6\uparrow\rangle$ undergoes a two-particle excitation $s_1$ at time $\tau_1$ replacing the occupied orbitals $\ket{\mathbf{k}_0\uparrow}, |\mathbf {k}_3\downarrow\rangle$ by 
%$|\mathbf{k}_2\uparrow\rangle, |\mathbf{k}_5\downarrow\rangle$. Two further excitations occur at $\tau_2$ and $\tau_3$.
%The states at the ``imaginary times'' $\tau = 0$ and $\tau = \beta$ coincide. In addition, the total spin projection is conserved at any time. All possible paths contribute to the partition function $Z$, Eq.~(\ref{eq:Z_expansion}).}
% \label{fig:sketch}
%\end{figure}
For CPIMC\cite{tim1,tim2}, instead of performing the trace over the density operator in the coordinate representation [see Eq.~(\ref{Z})], we trace over Slater determinants of the form
\begin{align}
|\{n\}\rangle=|n_1, n_2, \dots\rangle\;,
\end{align}
where, in case of the uniform electron gas, $n_i$ denotes the fermionic occupation number ($n_i\in\{0, 1\}$) of the $i$-th plane wave spin-orbital $|\mathbf{k}_i\sigma_i\rangle$. To obtain an expression for the partition function suitable for Metropolis Monte Carlo, we split the Hamiltonian into diagonal and off-diagonal parts, $\hat{H}=\hat{D}+\hat{Y}$ (with respect to the chosen plane wave basis, see Sec.~\ref{Sec:ms}), and explore a perturbation expansion of the density operator with respect to $\hat{Y}$:
\begin{align}
  e^{-\beta\hat{H}}&=e^{-\beta\hat{D}}\sum_{K=0}^\infty \int\limits_{0}^{\beta} d\tau_1 \int\limits_{\tau_1}^{\beta} d\tau_2 \ldots \int\limits_{\tau_{K-1}}^\beta d\tau_K\nonumber\\  
&\qquad\qquad(-1)^K\hat{Y}(\tau_K)\hat{Y}(\tau_{K-1})\cdot\ldots\cdot\hat{Y}(\tau_1)\;,
\label{pertExpansion}
\end{align}
with
$\hat{Y}(\tau)=e^{\tau\hat{D}}\hat{Y}e^{-\tau\hat{D}}$. In this representation the partition function becomes
\begin{align}
Z = &
%\begin{aligned}[t]
\sum_{K=0\atop (K \neq 1)}^{\infty} \sum_{\{n\}}
\sum_{s_1\ldots s_{K-1}}\,
\int\limits_{0}^{\beta} d\tau_1 \int\limits_{\tau_1}^{\beta} d\tau_2 \ldots \int\limits_{\tau_{K-1}}^\beta d\tau_K 
\label{eq:Z_expansion} \\\nonumber
& (-1)^K  
e^{-\sum\limits_{i=0}^{K} D_{\{n^{(i)}\}} \left(\tau_{i+1}-\tau_i\right) } 
\prod_{i=1}^{K} Y_{\{n^{(i)}\},\{n^{(i-1)}\} }(s_i)\;.
\end{align}
The matrix elements of the diagonal and off-diagonal operators are given by the Slater-Condon rules
 \begin{align}
 &D_{\{n^{(i)}\}} = \sum_l \mathbf{k}_l^2 n^{(i)}_{l} + \sum_{l<k}w^-_{lklk}n^{(i)}_{l}n^{(i)}_{k} \;,\label{eq:diagonal}\\
&Y_{ \{n^{(i)}\},\{n^{(i-1)}\} }(s_i) =w^-_{s_i}(-1)^{\alpha^{\phantom{-}}_{s_i}}\;,
\label{eq:off_diagonal} \\
&\alpha^{\phantom{-}}_{s_i} =\alpha^{(i)}_{pqrs}=\sum_{l=p}^{q-1}n^{(i-1)}_{l}+\sum_{l=r}^{s-1}n^{(i)}_{l}\;,
\end{align}
where the multi-index $s_i=(pqrs)$ defines the four orbitals in which $\{n^{(i)}\}$ and $\{n^{(i-1)}\}$ differ and we note that $p<q$ and $r<s$. As in standard PIMC, each contribution to the partition function~(\ref{eq:Z_expansion}) can be interpreted as a $\beta-$periodic path in imaginary time, but the path is now in Fock space instead of coordinate space. 
%An example for such a path is sketched in Fig.~\ref{fig:sketch}.
Evidently, the weight corresponding to any given path (second line of Eq.~(\ref{eq:Z_expansion})) can be positive or negative. Therefore, to apply the Metropolis algorithm, we have to proceed as explained in Sec.~\ref{sub:sign_problem} and use the modulus of the weight function as our probability density. In consequence, the CPIMC method is also afflicted with the FSP. However, as it turns out, the severity of the FSP as a function of the density parameter is complementary to that of standard PIMC, so that weakly interacting systems, which are the most challenging for PIMC, are easily tackled using CPIMC. For a detailed derivation of the CPIMC partition function and the Monte Carlo steps required to sample it see, e.g., Refs.~\onlinecite{tim1,tim2,groth,tim_prl}.

\subsection{\label{sub:dmqmc}Density matrix Quantum Monte Carlo}

Instead of sampling contributions to the partition function, as in path integral methods, DMQMC samples the (unnormalized) thermal density matrix directly by expanding it in a discrete basis of outer products of Slater determinants
\begin{equation}
    \hat{\rho} = \sum_{\{n\},\{n'\}} \rho_{\{n\},\{n'\}} |\{n\}\rangle\langle\{n'\}|,\label{eq:dmat}
\end{equation}
where $\rho_{\{n\},\{n'\}} = \langle \{n\} | e^{-\beta \hat{H}} | \{n'\} \rangle$. The density-matrix coefficients $\rho_{\{n\},\{n'\}}$ appearing in Eq.~(\ref{eq:dmat}) are found by simulating the evolution of the Bloch equation,
\begin{equation}
    \frac{d\hat{\rho}}{d\beta} = -\frac{1}{2}\left(\hat{\rho}\hat{H}+\hat{H}\hat{\rho}\right), \label{eq:itschr}
\end{equation}
which may be finite-differenced as
\begin{align}
    \rho_{\{n\},\{n'\}}(\beta+\Delta\beta) &= \rho_{\{n\},\{n'\}}(\beta) - \label{eq:sol_bloch}\\
                                           &   \qquad \Delta\beta \sum_{\{n''\}} \left[ \rho_{\{n\},\{n''\}}(\beta)H_{\{n''\},\{n'\}}\right.\nonumber\\
                                           &    \qquad\qquad \left. + H_{\{n\},\{n''\}}\rho_{\{n''\},\{n'\}}(\beta)\right]\nonumber.
\end{align}
The matrix elements of the Hamiltonian are as given as in Eqs.~(\ref{eq:diagonal}) and (\ref{eq:off_diagonal}).

    Following Booth and coworkers\cite{booth}, we note that Eq.~(\ref{eq:sol_bloch}) can be interpreted as a rate equation and can be solved by evolving a set of positive and negative walkers which stochastically undergo birth and death processes that, on average, reproduce the full solution. The rules governing the evolution of the walkers, as derived from Eq.~(\ref{eq:sol_bloch}), can be found elsewhere\cite{booth,blunt}. The form of $\hat{\rho}$ is known exactly at infinite temperature ($\beta=0$, $\hat{\rho} = \hat{1}$), providing an initial condition for Eq.~(\ref{eq:itschr}). For the electron gas, however, it turns out that simulating a differential equation that evolves a mean-field density matrix at inverse temperature $\beta$ to the exact density matrix at inverse temperature $\beta$ is much more efficient than solving Eq.~(\ref{eq:itschr}), an insight that leads to the `interaction picture' version of DMQMC\cite{malone2,malone} used throughout this work.

The sign problem manifests itself in DMQMC as an exponential growth in the number of walkers required for the sampled density matrix to emerge from the statistical noise\cite{booth,spencer,kolodrubetz,shepherd3}. Working in a discrete Hilbert space helps to reduce the noise by ensuring a more efficient cancellation of positive and negative contributions, enabling larger systems and lower temperatures to be treated than would otherwise be possible. Nevertheless, at some point the walker numbers required become overwhelming and approximations are needed. Recently, we have applied the initiator approximation\cite{cleland,shepherd,shepherd2} to DMQMC ($i-$DMQMC). In principle, at least, this allows a systematic approach to the exact result with increasing walker number. More details on the use of the initiator approximation in DMQMC and its limitations can be found in Ref.~\onlinecite{malone2}.

\section{\label{sec:finite}Simulation results for the finite system}

\begin{figure}[]
 \centering
\includegraphics[width=0.43\textwidth]{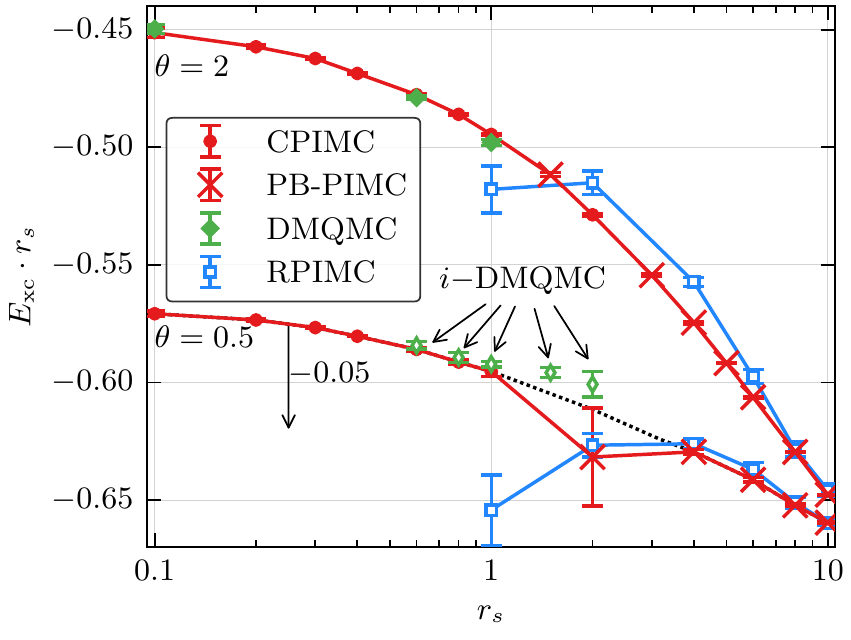}
 \caption{\label{fig:all_methods} 
 Exchange-correlation energy of $N=33$ spin-polarized electrons as a function of the density parameter $r_s$ for two isotherms. Shown are results from CPIMC and PB-PIMC taken from Ref.~\onlinecite{groth}, restricted PIMC from Ref.~\onlinecite{brown}, and DMQMC from Ref.~\onlinecite{malone2}. For $\theta=0.5$, all data has been shifted by $0.05$ Hartree. In the case of DMQMC, the initiator approximation is used.
}
\end{figure}
The first step towards obtaining QMC results for the warm dense electron gas in the thermodynamic limit is to carry out accurate simulations of a finite model system. In Fig.~\ref{fig:all_methods}, we compare results for the density dependence of the exchange correlation energy $E_{xc}$ of the UEG for $N=33$ spin-polarized electrons and two different temperatures. The first results, shown as blue squares, were obtained with RPIMC \cite{brown} for $r_s\geq1$. Subsequently, Groth, Dornheim and co-workers\cite{dornheim2,groth} showed that the combination of PB-PIMC (red crosses) and CPIMC (red circles) allows for an accurate description of this system for $\theta\geq0.5$. In addition, it was revealed that RPIMC is afflicted with a systematic nodal error for densities greater than the relatively low value at which $r_s=6$. Nevertheless, the FSP precludes the use of PB-PIMC at lower temperatures and, even at $\theta=0.5$ and $r_s=2$, the statistical uncertainty becomes large.
    The range of applicability of DMQMC is similar to that of CPIMC and the DMQMC results (green diamonds) fully confirm the CPIMC results\cite{malone,malone2}. Further, the introduction of the initiator approximation (i-DMQMC) has made it possible to obtain results up to $r_s=2$ for $\theta=0.5$. Although i-DMQMC is, in principle, systematically improvable and controlled, the results suggest that the initiator approximation may introduce a small systematic shift at higher densities.

In summary, the recent progress in fermionic QMC methods has resulted in a consensus regarding the finite-$N$ UEG for temperatures $\theta\geq0.5$. 
However, there remains a gap at $r_s\approx2-6$ and $\theta<0.5$ where, at the moment, no reliable data are available.

\section{\label{sec:finite_size}Finite size corrections }
In this section, we describe in detail the finite-size correction scheme introduced in Ref.~\onlinecite{dornheim_prl} and subsequently present detailed results for two elucidating examples.
\subsection{\label{sub:fs_theory}Theory}

As introduced above (see Eq.~(\ref{Hcoord}) in Sec.~\ref{cor}), the potential energy of the finite simulation cell is defined as the interaction energy of the $N$ electrons with each other, the infinite periodic array of images, and the uniform positive background.
To estimate the finite-size effects, it is more convenient to express the potential energy in $k$-space. For the finite simulation cell of $N$ electrons, the expression obtained is a sum over the discrete reciprocal lattice vectors $\mathbf{G}$:
\begin{eqnarray}
\frac{V_N}{N}=
\label{EQ:V_N} \frac{1}{2\Omega}\sum_{\mathbf{G}\ne\mathbf{0}}\left[S_N(\mathbf{G})-1\right]\frac{4\pi}{G^2}+\frac{\xi_\textnormal{M}}{2} \ ,
\end{eqnarray}
where $S(\mathbf{k})$ is the static structure factor.
In the limit as the system size tends to infinity  and $\xi_{\textnormal{M}}\rightarrow 0$, this yields the integral
\begin{eqnarray}
v =
\label{EQ:v} \frac{1}{2}\int_{k<\infty}\frac{\textnormal{d}\mathbf{k}}{(2\pi)^3} \left[S(k)-1\right]\frac{4\pi}{k^2} \ .
\end{eqnarray}
Combining Eqs.~(\ref{EQ:V_N}) and (\ref{EQ:v}) yields the finite-size error for a given QMC simulation:
\begin{eqnarray}
& &\frac{\Delta V_N}{N}[S(k),S_N(k)] = v - \frac{V_N}{N}  \label{EQ:delta_V} \\ \nonumber
 & &=  \underbrace{ \frac{1}{2}\int_{k<\infty}\frac{\textnormal{d}\mathbf{k}}{(2\pi)^3} \left[S(k)-1\right]\frac{4\pi}{k^2} }_{v} \notag \\  
 & & - \underbrace{ \left(
 \frac{1}{2L^3}\sum_{\mathbf{G}\ne\mathbf{0}}\left[S_N(\mathbf{G})-1\right]\frac{4\pi}{G^2}+\frac{\xi_\textnormal{M}}{2}\right) }_{V_N/N} \label{EQ:delta_V2} .
\end{eqnarray}

The task at hand is to find an accurate estimate of the finite-size error from Eq.~(\ref{EQ:delta_V}), which, when added to the QMC result for $V_N/N$, gives the potential energy in the thermodynamic limit.
As a first step, we note that the Madelung constant may be approximated by\cite{drummond}
\begin{equation}
\quad\xi_\textnormal{M}\approx \frac{1}{L^3} \sum_{\mathbf{G}\neq \mathbf{0}} \frac{4\pi}{G^2} e^{-\epsilon G^2}
- \frac{1}{(2\pi)^3}\int_{k < \infty}\textnormal{d}\mathbf{k}\frac{4\pi}{k^2}e^{-\epsilon k^2}\ ,
\end{equation}
an expression that becomes exact in the limit as $\epsilon \rightarrow 0$.
The Madelung term thus cancels the minus unity contributions to both the sum and the integral in Eq.~(\ref{EQ:delta_V2}).

The remaining two possible sources of the finite-size error in Eq.~(\ref{EQ:delta_V}) are: (i) The substitution of the static structure factor of the infinite system $S(k)$ by its finite-size equivalent $S_N(k)$; and (ii) the approximation of the continuous integral by a discrete sum, resulting in a discretization error. As we will show in Sec.~\ref{sub:fs_results}, $S_N(k)$ exhibits a remarkably fast convergence with system size, which leaves explanation (ii).
In particular, about a decade ago, Chiesa \textit{et al.}~\cite{chiesa} suggested that the main contribution to Eq.~(\ref{EQ:delta_V}) stems from the $\mathbf{G}=0$ term that is completely missing from the discrete sum.
To remedy this shortcoming, they made use of the random phase approximation (RPA) for the structure factor, which becomes exact in the limit $k\to0$. The leading term in the expansion of $S^{\rm RPA}(k)$ around $k=0$ is\cite{rpa0}
\begin{equation}
\label{eq:RPA_zero}S^{\rm RPA}_0(k) = \frac{k^2}{2\omega_p}\textnormal{coth}\left( \frac{\beta\omega_p}{2} \right) ,
\end{equation}
with $\omega_p=\sqrt{3/r_s^3}$ being the plasma frequency. The finite-$T$ generalization of the Chiesa \textit{et al.}~FSC, hereafter called the BCDC-FSC, is \cite{brown,dornheim_prl}:
\begin{eqnarray}\nonumber
\Delta V_\textnormal{BCDC}(N) &=& \lim_{k\rightarrow 0} 
\frac{S_0^{\textnormal{RPA}}(k)4 \pi}{2 L^3 k^2} \\
&=& \frac{\omega_p}{4N} \textnormal{coth}\left( \frac{\beta\omega_p}{2} \right) . \label{eq:BCDC}
\end{eqnarray}
Eq.~(\ref{eq:BCDC}) would be sufficient if (i) $S_0^\textnormal{RPA}(k)$ were accurate for $k\lesssim 2\pi/L$, and (ii) all contributions to Eq.~(\ref{EQ:delta_V}) beyond the $\mathbf{G}=\mathbf{ 0 }$ term were negligible.
As is shown in Sec.~\ref{sub:fs_results}, both conditions are strongly violated in parts of the warm dense regime.

To overcome the deficiencies of Eq.~(\ref{eq:BCDC}), we need a continuous model function $S_\textnormal{model}(k)$ to accurately estimate the discretization error from Eq.~(\ref{EQ:delta_V2}):
\begin{eqnarray}
\label{eq:improved}\Delta V_N[S_\textnormal{model}(k)] = \frac{\Delta V_N}{N}[S_\textnormal{model}(k),S_\textnormal{model}(k)] \ .
\end{eqnarray}
A natural choice would be to combine the QMC results for $k\geq k_\textnormal{min}$, which include all short-ranged correlations and exchange effects, with the STLS structure factor for smaller $k$, which is exact as $k\to0$ and incorporates the long-ranged behavior that cannot be reproduced using QMC due to the limited size of the simulation cell. However, as we showed in Ref.~\onlinecite{dornheim_prl}, a simpler approach using $S_\textnormal{STLS}(k)$ [or the full RPA structure factor $S_\textnormal{RPA}(k)$] for all $k$ is sufficient to accurately estimate the discretization error.

\subsection{\label{sub:fs_results}Results}

%\begin{figure}[]
% \centering
%\includegraphics[width=0.44\textwidth]{rs_Potential.pdf}\vspace*{-0.4cm}
%\includegraphics[width=0.44\textwidth]{theta_Potential.pdf}\vspace*{-0.25cm}
% \caption{\label{fig:fs_comparison} 
%Relative finite-size correction as a function of density and temperature -
%Top figure: $r_s$-dependence for $\theta=2$.\\
%Bottom figure: $\theta$-dependence for $r_s=1$.
%}
%\end{figure}

\subsubsection{\label{subsub:N}Particle number dependence}

\begin{figure*}[]
 \centering

 \begin{minipage}{0.47\textwidth}
 \vspace*{-0.06cm}
 \hspace*{0.16cm}\includegraphics[width=.94\textwidth]{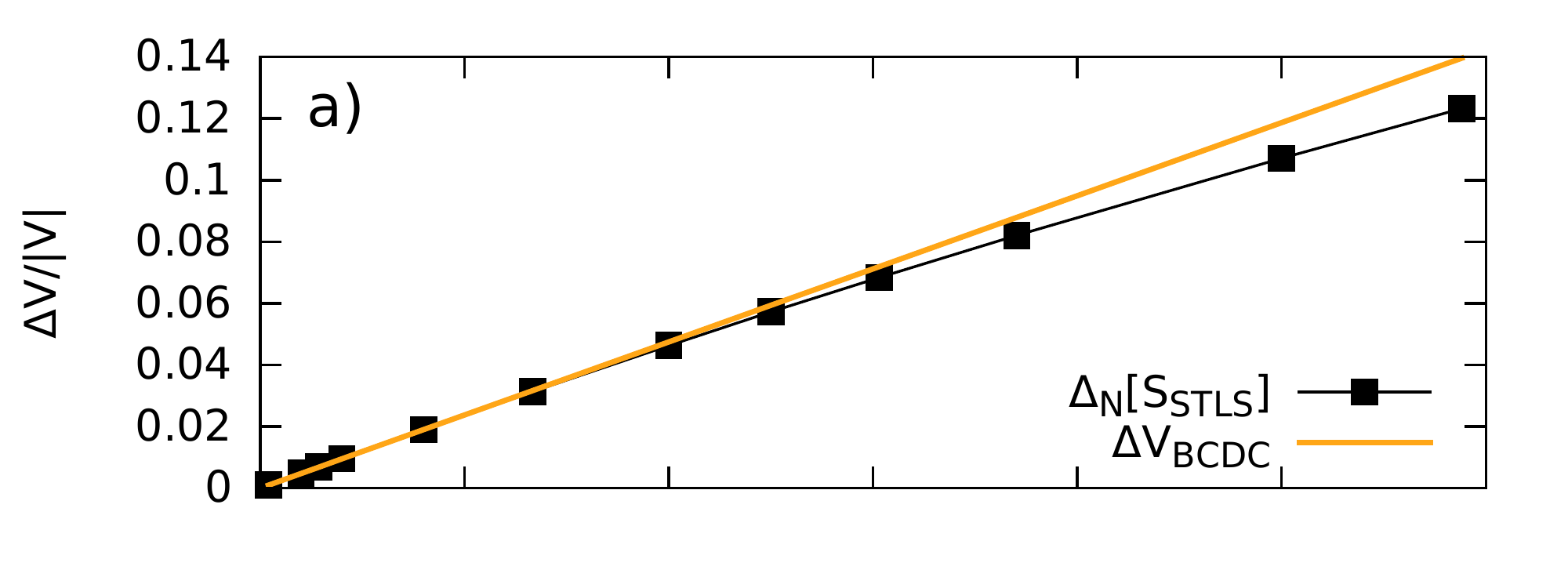}\vspace*{-0.69cm}\\
 \hspace*{0.16cm}\includegraphics[width=.94\textwidth]{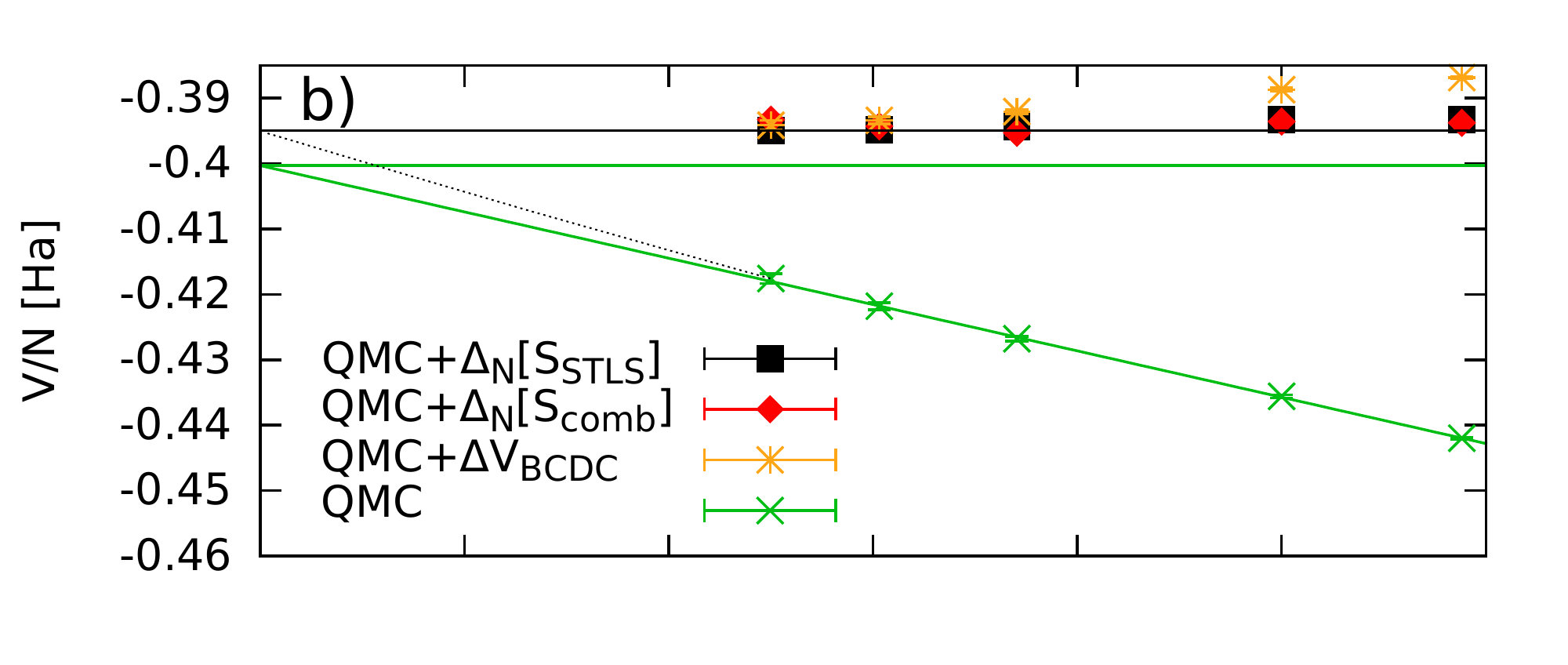}\vspace*{-0.74cm}\\
 \hspace*{-0.04cm}\includegraphics[width=.965\textwidth]{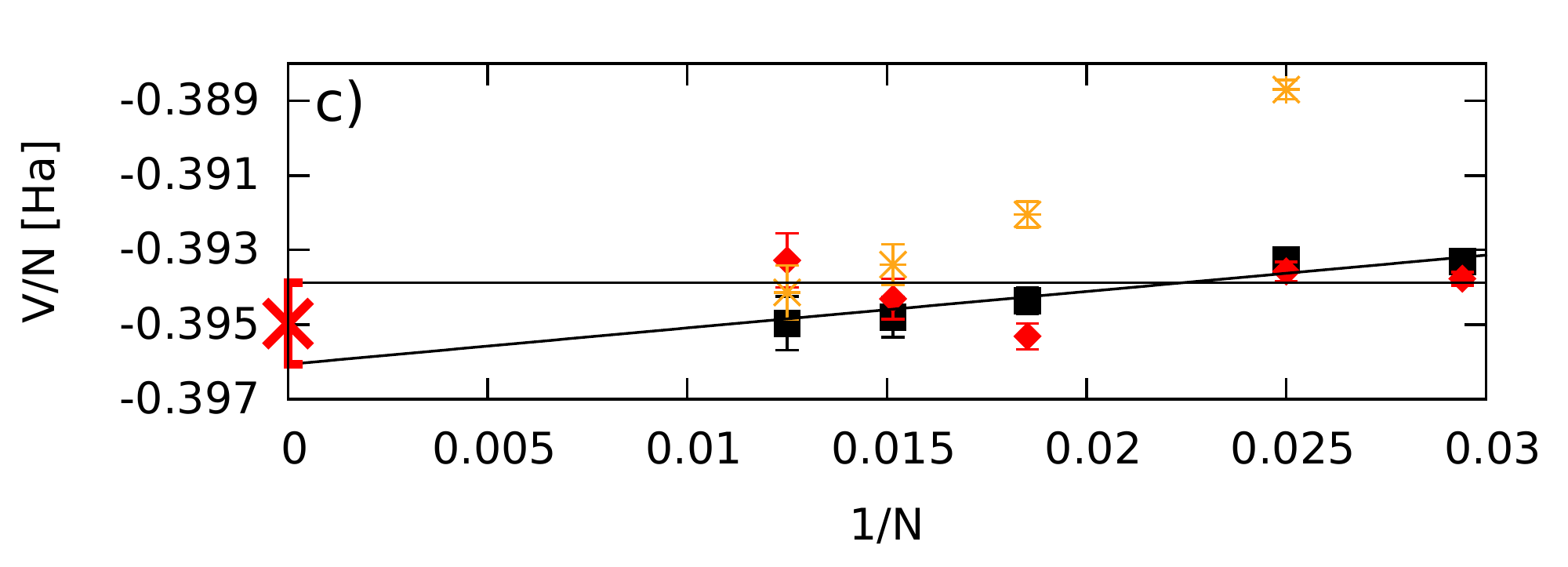}

\end{minipage}\begin{minipage}{0.47\textwidth}

\includegraphics[width=.94\textwidth]{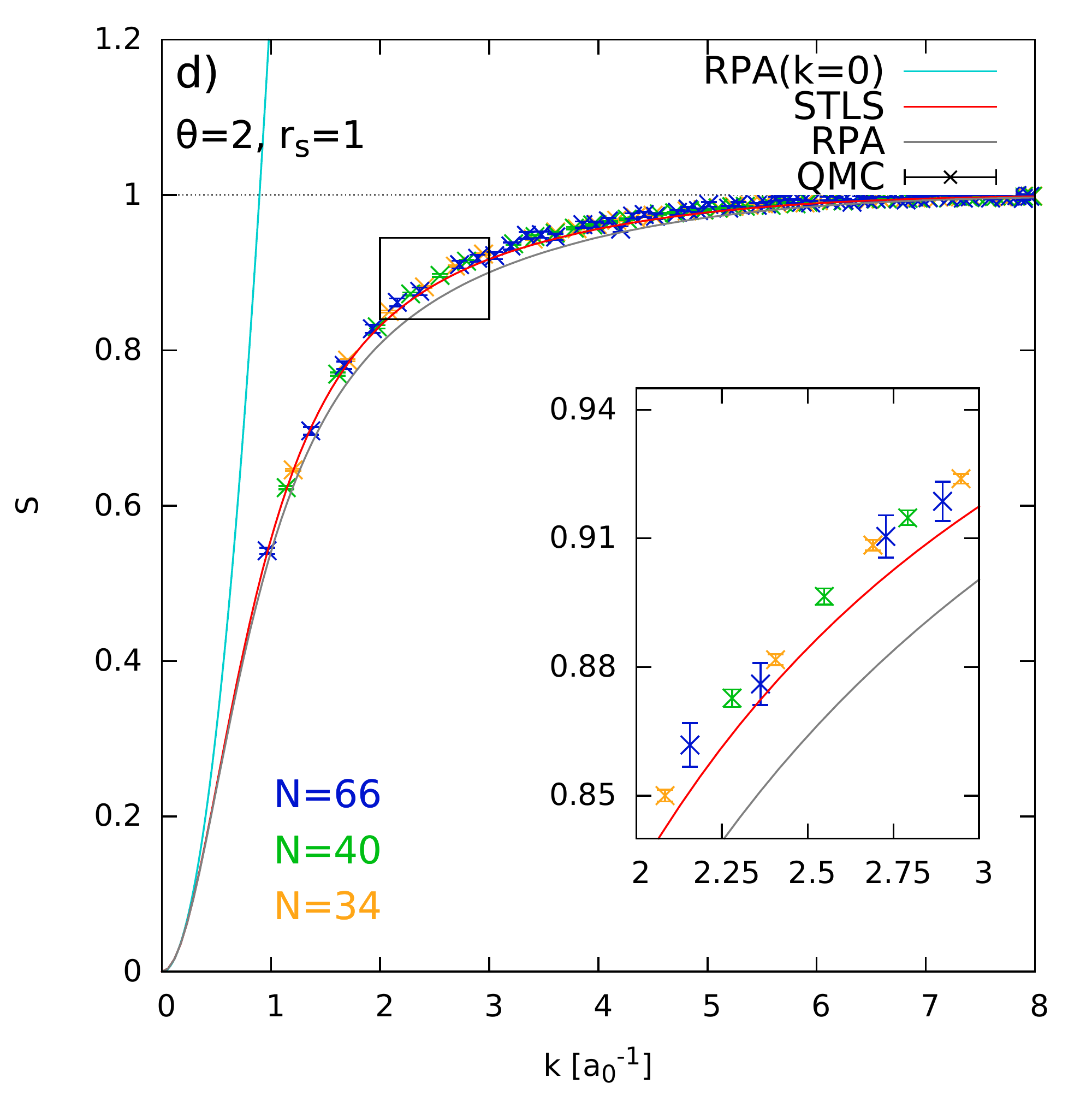}

\end{minipage}

 \caption{\label{fig:fs_theta2_rs1} Finite-size correction for the UEG at $\theta=2$ and $r_s=1$: a) $N$ dependence of the FSCs; b) potential energy per particle, $V/N$; the dotted grey line corresponds to the TDL value where the $\Delta_N[S_\textnormal{STLS}]$ had been substracted; c) extrapolation of the residual finite-size error; and d) corresponding static structure factors $S(k)$ from QMC (for $N=34,40,66$), STLS, RPA, and the RPA expansion around $k=0$, Eq.~(\ref{eq:RPA_zero}).
}
\end{figure*}
To illustrate the application of the different FSCs, Fig.~\ref{fig:fs_theta2_rs1} shows results for the unpolarized UEG at $\theta=2$ and $r_s=1$. The green crosses in panel b) correspond to the raw, uncorrected QMC results that, clearly, are not converged with system size $N$. The raw data points appear to fall onto a straight line when plotted as a function of $1/N$. This agrees with the BCDC-FSC formula, Eq.~(\ref{eq:BCDC}), which also predicts a $1/N$ behavior, and suggests the use of a linear extrapolation (the green line). However, while the linear fit does indeed exhibit good agreement with the QMC results, the computed slope does not match Eq.~(\ref{eq:BCDC}). Further, the points that have been obtained by adding $\Delta V_\textnormal{BCDC}$ to the QMC results, i.e., the yellow asterisks, do not fall onto a horizontal line and do not agree with the prediction of the linear extrapolation (see the horizontal green line).
To resolve this peculiar situation, we compute the improved finite-size correction [Eq.~(\ref{eq:improved})] using both the static structure factor from STLS ($S_\textnormal{STLS}$) and the combination of STLS with the QMC data ($S_\textnormal{comb}$) as input.
The resulting corrected potential energies are shown as black squares and red diamonds, respectively, and appear to exhibit almost no remaining dependence on system size. In panel c) we show a segment of the corrected data, magnified in the vertical direction. Any residual finite-size errors [due to the QMC data for $S(k)$ not being converged with respect to $N$, see panel d)] can hardly be resolved within the statistical uncertainty and are removed by an additional extrapolation. In particular, to compute the final result for $V/N$ in the thermodynamic limit, we obtain a lower bound via a linear extrapolation of the corrected data (using $S_\textnormal{STLS}$) and an upper bound by performing a horizontal fit to the last few points, all of which are converged to within the error bars.
% \textcolor{red}{[WMCF: In panel c), it is not clear whether the grey and black lines are obtained by fitting the red points or the black points. The black line appears to be a fit to the black points, but the grey line could fit either and the use of a red error bar at $1/N=0$ further confuses the issue. Also, the horizontal green line in panel b) needs explanation.]} 
The dotted grey line in panel b), which connects to the extrapolated result, shows clearly that the results of this procedure deviate from the results of a naive linear extrapolation.

Finally, in panel d) of Fig.~\ref{fig:fs_theta2_rs1}, we show results for the static structure factor $S(k)$ for the same system. As explained in Sec.~\ref{sub:fs_theory}, momentum quantization limits the QMC results to discrete $k$ values above a minimum value $k_\textnormal{min}=2\pi/L$. Nevertheless, the $N$ dependence of the $k$ grid is the only apparent change of the QMC results for $S(k)$ with system size and no difference between the results for the three particle numbers studied can be resolved within the statistical uncertainty (see also the magnified segment in the inset). 
The STLS curve (red) is known to be exact in the limit $k\to0$ and smoothly connects to the QMC data, although for larger $k$ there appears an almost constant shift. The full RPA curve (grey) exhibits a similar behavior, albeit deviating more significantly at intermediate $k$.
Finally, the RPA expansion around $k=0$ [Eq.~(\ref{eq:RPA_zero}), light blue] only agrees with the STLS and full RPA curves at very small $k$ and does not connect to the QMC data even for the largest system size simulated.

\begin{figure*}[]
 \centering

 \begin{minipage}{0.47\textwidth}
 \vspace*{-0.06cm}
 \hspace*{0.16cm}\includegraphics[width=.94\textwidth]{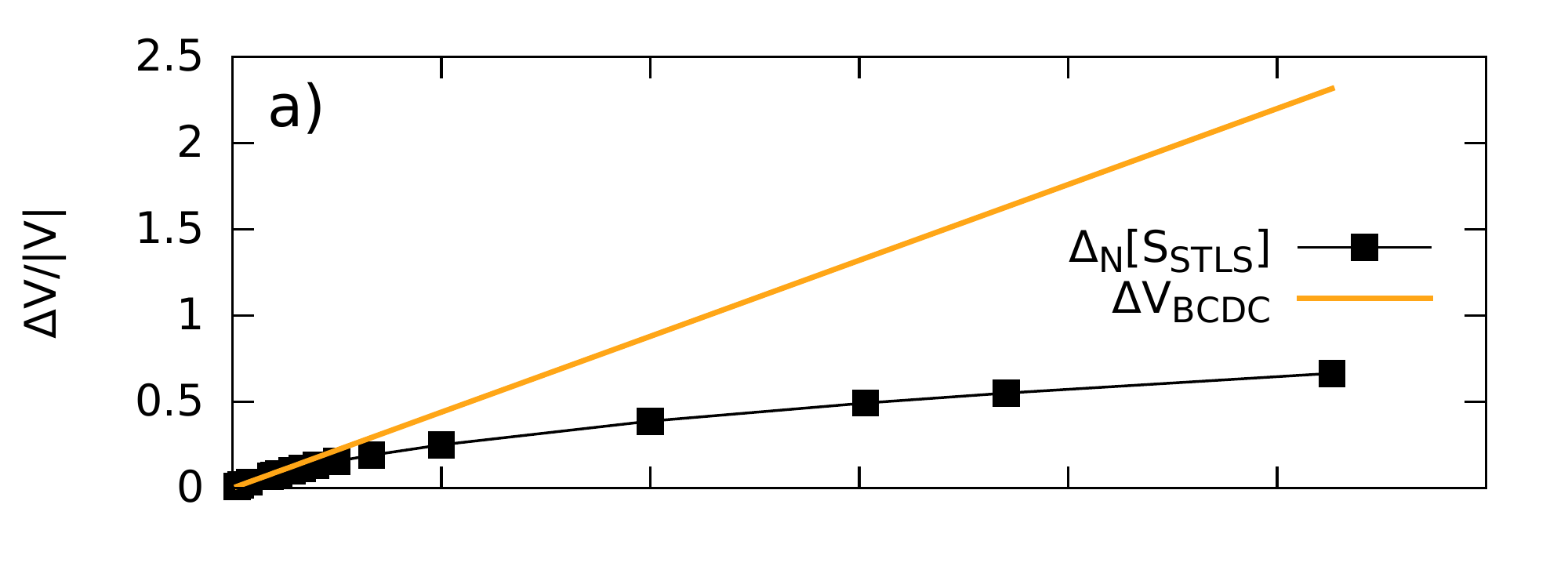}\vspace*{-0.69cm}\\
 \hspace*{0.16cm}\includegraphics[width=.94\textwidth]{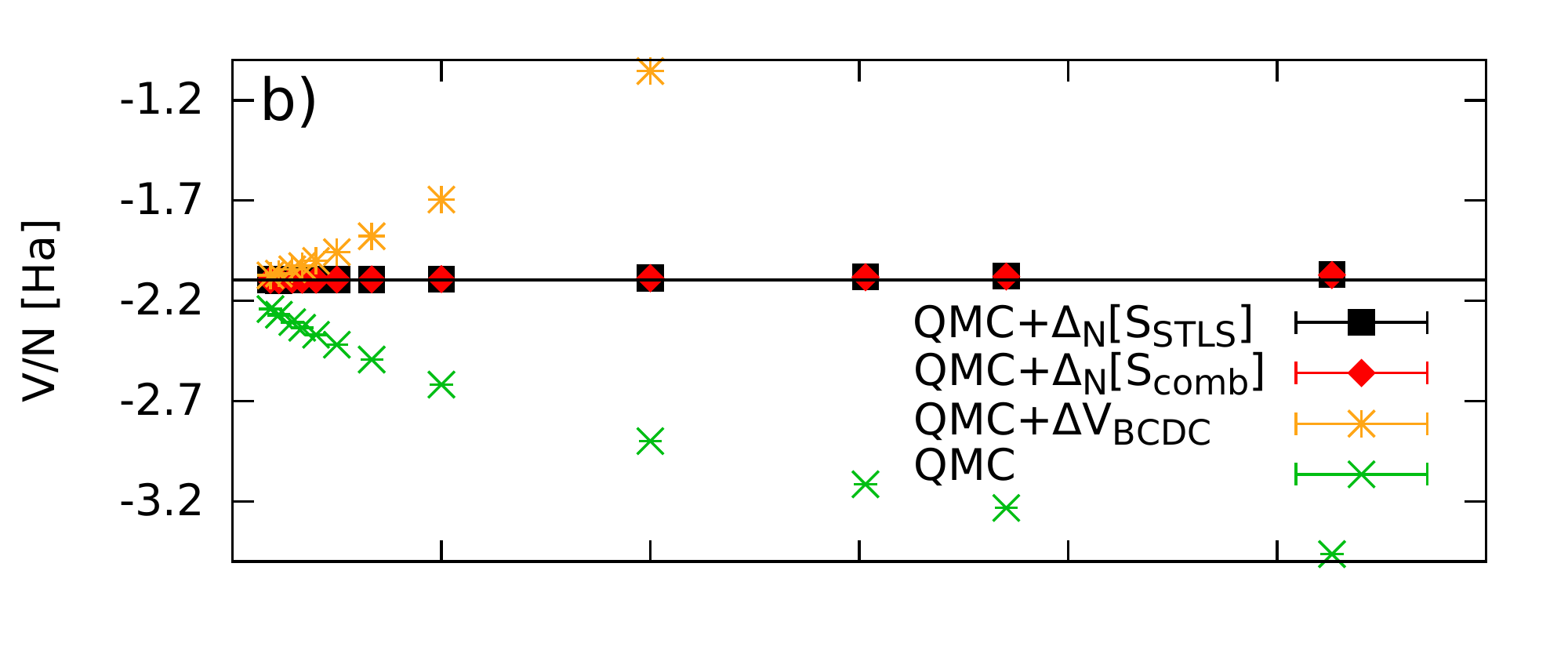}\vspace*{-0.74cm}\\
 \hspace*{-0.18cm}\includegraphics[width=.985\textwidth]{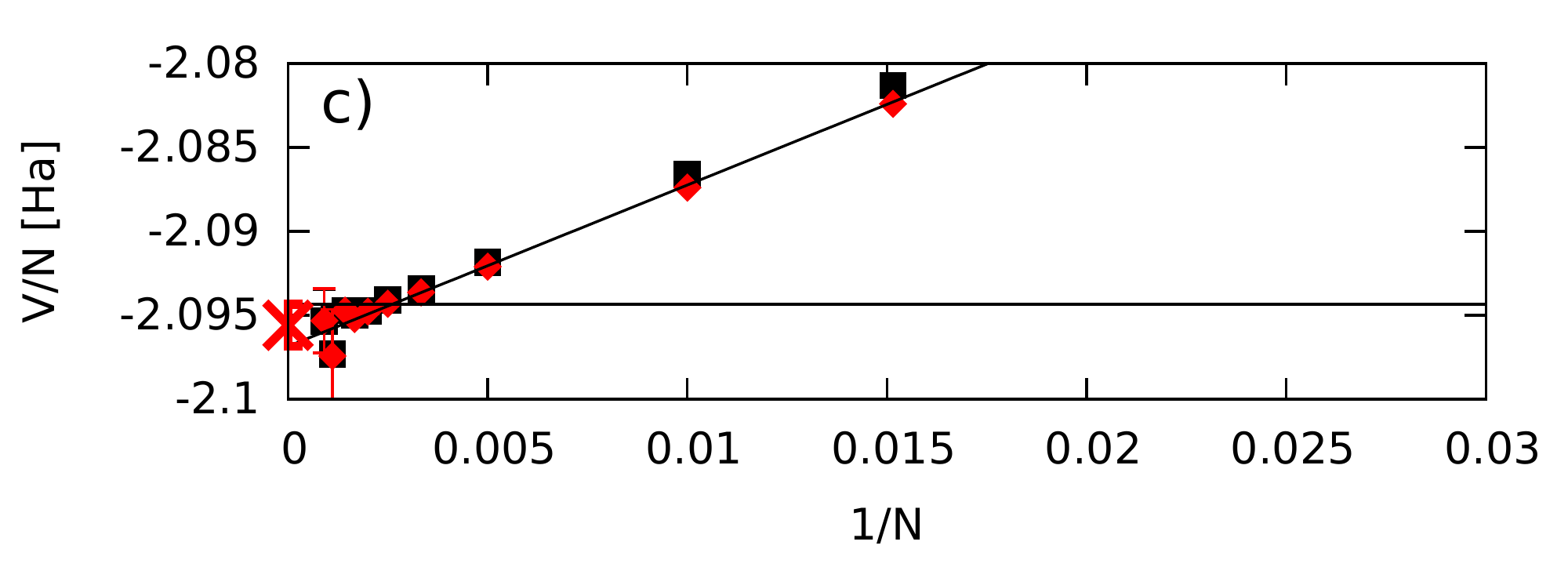}

\end{minipage}\begin{minipage}{0.47\textwidth}

\includegraphics[width=.94\textwidth]{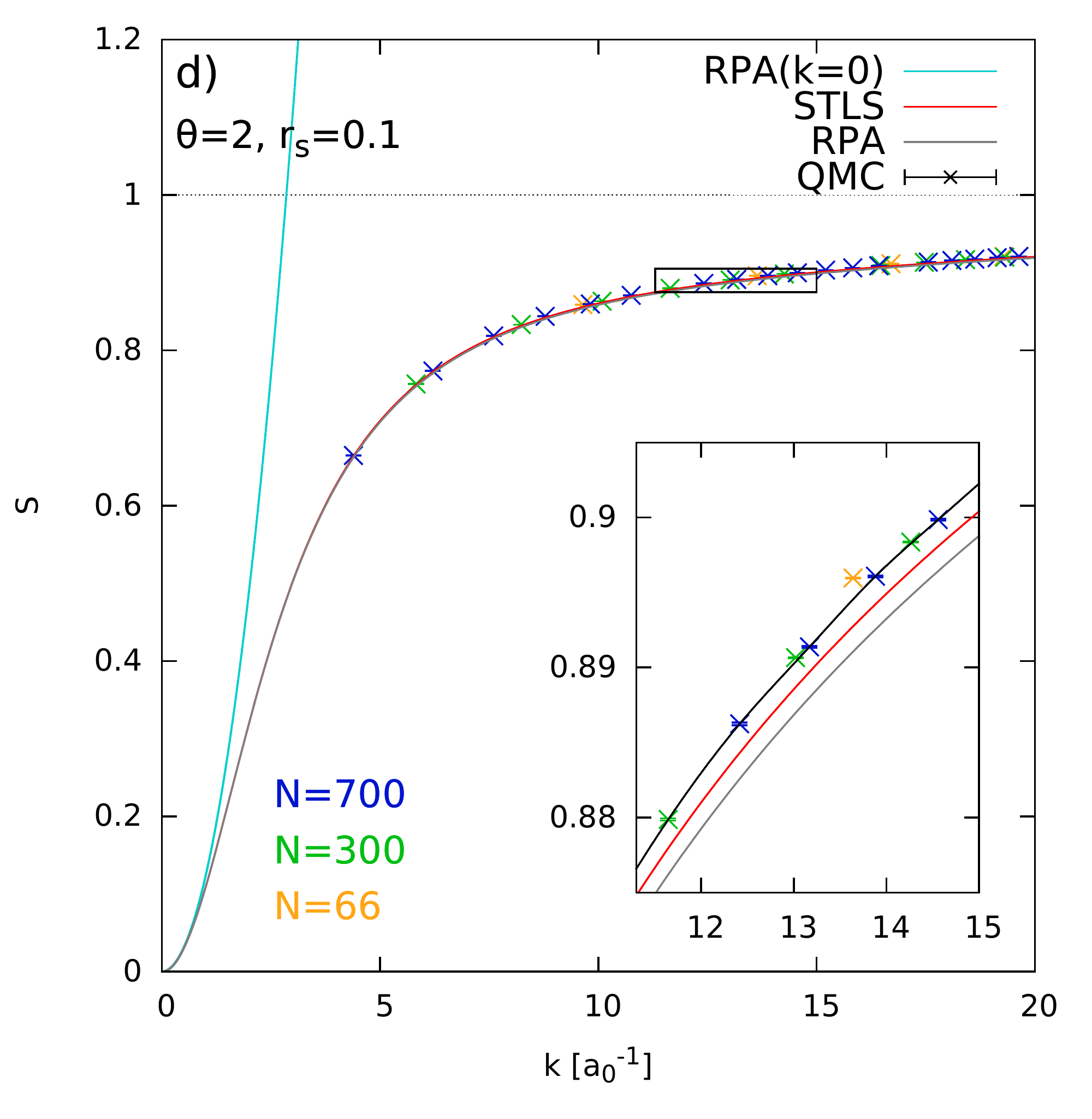}

\end{minipage}

\caption{\label{fig:fs_theta2_rs0p1} Finite-size correction for the UEG at $\theta=2$ and $r_s=0.1$: a) $N$ dependence of the FSCs; b) potential energy per particle, $V/N$; c) extrapolation of the residual finite-size error; and d) corresponding static structure factors $S(k)$ from QMC (for $N=66, 300, 700$), STLS, RPA, and the RPA expansion around $k=0$, Eq.~(\ref{eq:RPA_zero}).
}
\end{figure*}

To further stress the importance of our improved finite-size correction scheme, Fig.~\ref{fig:fs_theta2_rs0p1} shows results again for $\theta=2$ but at higher density, $r_s=0.1$. In this regime, the CPIMC approach (and also DMQMC) is clearly superior to PB-PIMC and simulations of $N=700$ unpolarized electrons in $N_b=189234$ basis functions are feasible. Due to the high density, the finite-size errors are drastically increased compared to the previous case and exceed $50\%$ for $N=38$ particles [see panels a) and b)]. Further, we note that the BCDC-FSC is completely inappropriate for the $N$ values considered, as the yellow asterisks are clearly not converged and differ even more strongly from the correct TDL than the raw uncorrected QMC data. 

Our improved FSC, on the other hand, reduces the finite-size errors by two orders of magnitude (both with $S_\textnormal{STLS}$ and $S_\textnormal{comb}$) and approaches Eq.~(\ref{eq:BCDC}) only in the limit of very large systems [$N\gtrsim10^4$; see panel a)]. The small residual error is again extrapolated, as shown in~panel c).

Finally, we show the corresponding static static structure factors in panel d).
The RPA expansion is again insufficient to model the QMC data, while the full RPA and STLS curves smoothly connect to the latter.

%\begin{figure}[]
% \centering
%\includegraphics[width=0.44\textwidth]{isotherm_theta2.pdf}\vspace*{-0.25cm}
% \caption{\label{fig:isotherm_theta2} 
%Potential energy per particle of the uniform electron gas at $\theta=2$ as a function of $r_s$. Shown are uncorrected %QMC results for $N=66$ particles (see Ref.~\cite{dornheim3}) and the finite-size corrected data from %Ref.~\cite{dornheim_prl}.
%}
%\end{figure}

 \subsubsection{Comparison to other methods\label{ss_compare}}
 \begin{figure}[]
 \centering
 \hspace*{0.16cm}\includegraphics[width=0.43\textwidth]{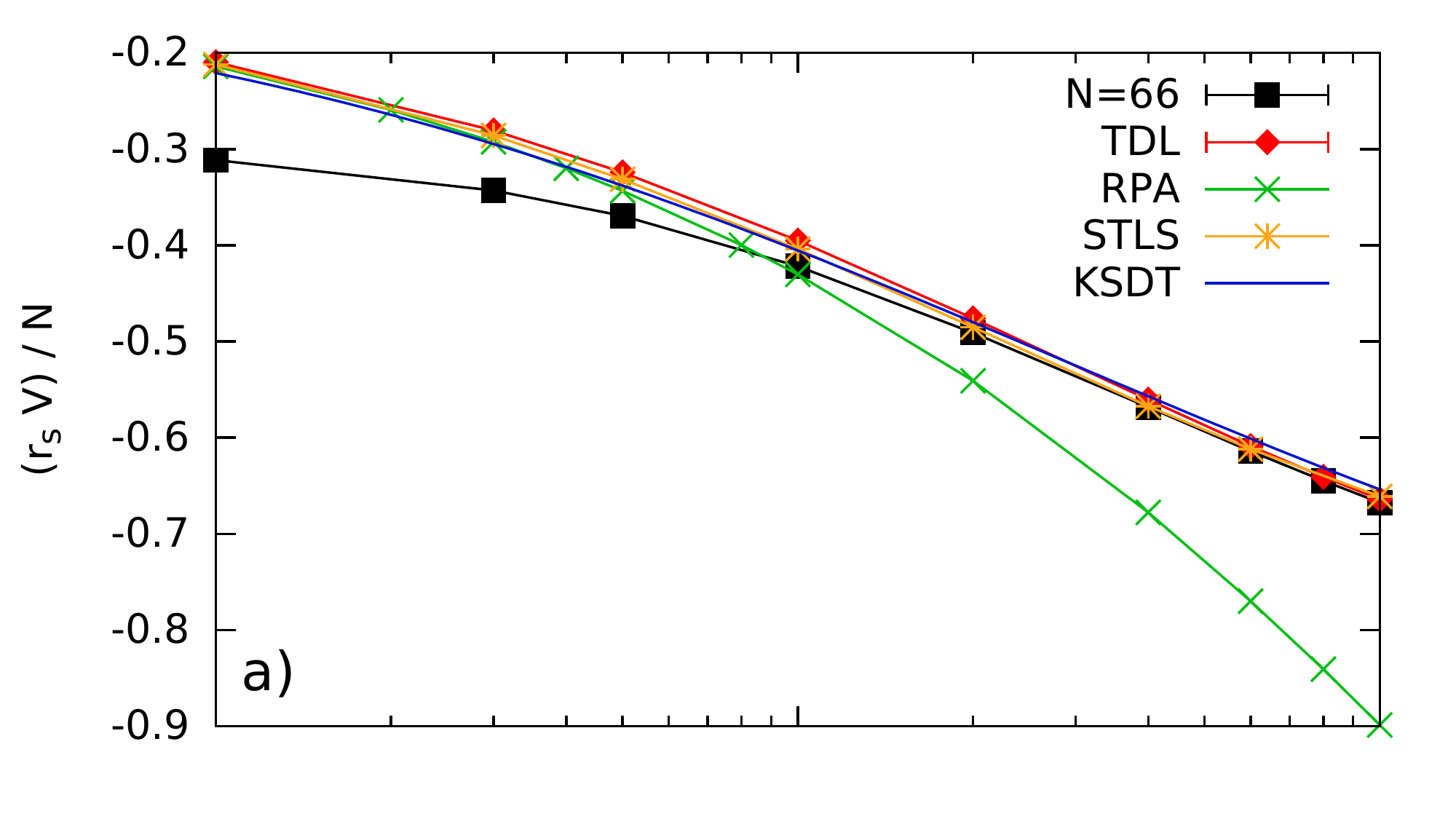}\vspace*{-0.72cm}
 \includegraphics[width=0.44\textwidth]{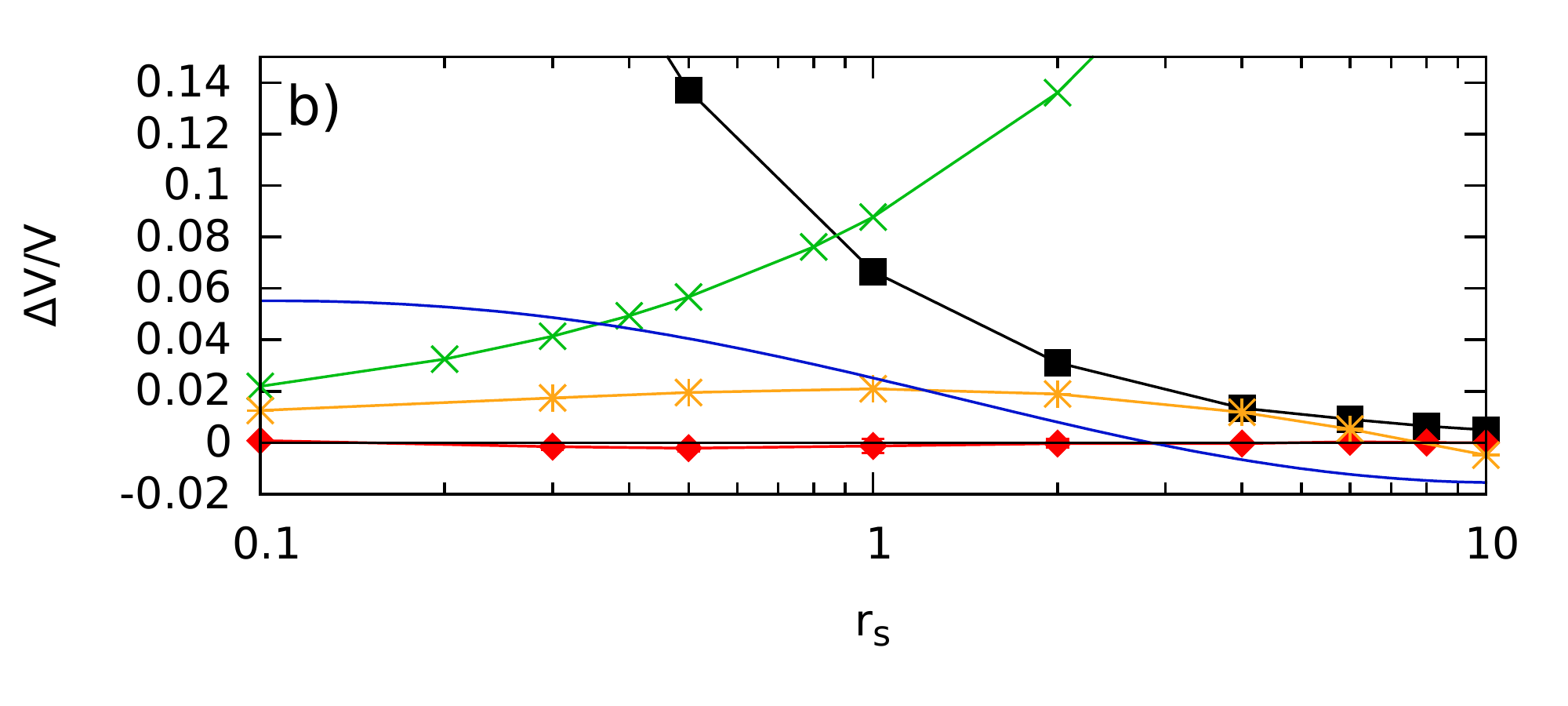}\vspace*{-0.52cm}
 \caption{\label{fig:isotherm_theta2} 
Potential energy per particle of the uniform electron gas at $\theta=2$--simulations versus analytical models. Squares: QMC results for $N=66$ particles~\cite{dornheim3}, (red) rhombs: finite-size corrected QMC data (TDL)~\cite{dornheim_prl}, green (yellow) curves: RPA (STLS) data~\cite{stls2}, blue: results of the parametrization of Ref.~\cite{karasiev} (KSDT). Lower Fig.: relative deviations of all curves from the fit to the thermodynamic QMC results.
}
\end{figure}

To conclude this section, we use our finite-size corrected QMC data for the unpolarized UEG to analyze the accuracy of various other methods that are commonly used.
In Fig.~\ref{fig:isotherm_theta2} a), the potential energy per particle, $V/N$, is shown as a function of $r_s$ for the isotherm with $\theta=2$.
Although all four depicted curves exhibit qualitatively similar behavior, there are significant deviations between them [see panel b), where we show the relative deviations from a fit to the QMC data in the TDL].
Let us start with the QMC results: the black squares correspond to the uncorrected raw QMC data for $N=66$ particles (see Ref.~\onlinecite{dornheim3}) and the red diamonds to the finite-size corrected data from Ref.~\onlinecite{dornheim_prl}.
As expected, the finite-size effects drastically increase with density from $|\Delta V|/V\approx 1\%$, at $r_s=10$, to $|\Delta V|/V \geq 50\%$, at $r_s=0.1$. This again illustrates the paramount importance of accurate finite-size corrections for QMC simulations in the warm dense matter regime.
The RPA calculation (green curve) is accurate at high density and weak coupling. However, with increasing $r_s$ the accuracy quickly deteriorates and, already at moderate coupling, $r_s=1$, the systematic error is of the order of $10\%$.
The yellow asterisks show the SLTS result which agrees well with the simulations (the systematic error does not exceed $3\%$) over the entire $r_s$-range considered, i.e., up to $r_s=10$.
Finally, the blue curve has been obtained from the recent parametrization of $f_{xc}$ by Karasiev \textit{et al.}\cite{karasiev} (KSDT), for which RPIMC data have been used as an input. While there is a reasonable agreement with our new data for $r_s\gtrsim 1$ (with $|\Delta V|/V\sim2\%$), there are significant deviations at smaller $r_s$, which only vanish for $r_s<10^{-4}$.

\section{\label{sec:discussion}Summary and open questions}
Let us summarize the status of \textit{ab initio} thermodynamic data for the uniform electron gas at finite temperature. The present paper has given an overview of recent progress in {\em ab initio} finite temperature QMC simulations that avoid any additional simplifications such as fixed nodes. While these simulations do not ``solve'' the fermion sign problem, they provide a reasonable and efficient way how to {\em avoid it}, in many practically relevant situations, by combining simulations that use different representations of the quantum many-body state: the coordinate representation (direct PIMC and PB-PIMC) and Fock states (CPIMC, DMQMC). With this it is now possible to obtain highly accurate results for up to $N\sim 100$ particles in the entire density range and for temperatures $\theta \gtrsim 0.5$. As a second step we demonstrated that these comparatively small simulation sizes are sufficient to predict results for the macroscopic uniform electron gas {\em not significantly loosing accuracy}~\cite{dornheim_prl}. This unexpected result is a consequence of a new highly accurate finite-size correction that was derived by invoking STLS results for the static structure factor.

With this procedure it is now possible to obtain thermodynamic data for the uniform electron gas with an accuracy on the order of $0.1\%$. Even though pure electron gas results cannot be directly compared to   warm dense matter experiments, they are of high value to benchmark and improve additional theoretical approaches. Most importantly, this concerns finite-temperature versions of density functional theory (such as orbital-free DFT) which is the standard tool to model realistic materials and which will benefit from our results for the exchange-correlation free energy. Furthermore, we have also presented a few comparisons with earlier models such as RPA, Vahista-Singwi, STLS or the recent fit of Karasiev {\em et al.} (KSDT) the accuracy and errors of which can now be unambiguously quantified. We found that among the tested models, the STLS is the most accurate one. We wish to underline that even though exchange-correlation effects are often small compared to the kinetic energy, their accurate treatment is important to capture the properties of real materials, see e.g. \cite{perdew_13}

In the following we summarize the open questions and outline future research directions.
\begin{enumerate}
  \item Construction of an improved fit for the exchange-correlation free energy due to their key relevance as input for finite-temperature DFT. Such fits are straightforwardly generated from the current results but require a substantial extension of the simulations to arbitrary spin polarization. This work is currently in progress.
  \item The presently available accurate data are limited to temperatures above half the Fermi energy, as a consequence of the fermion sign problem. A major challenge will be to advance to lower temperatures, $\Theta < 0.5$ and to reliably connect the results to the known ground state data. This requires substantial new developments in the area of the three quantum Monte Carlo methods presented in this paper (CPIMC, PB-PIMC and DMQMC) and new ideas how to combine them. Another idea could be to derive simplified versions of these methods that treat the FSP more efficiently but still have acceptable accuracy.
  \item
  The present {\em ab initio} results allow for an entirely new view on previous theoretical models. For the first time, a clear judgement about the accuracy becomes possible which more clearly maps out the sphere of applicability of the various approaches, e.g.~\cite{4ebeling}. Moreover, the availability of our data will allow for improvements of many of these approaches via adjustment of the relevant parameters to the QMC data. This could yield, e.g., improved static structure factors, dielectric functions or 
  local field correlations. 
  \item
  Similarly, our data may also help to improve alternative quantum Monte Carlo concepts. In particular, this concerns the nodes for Restricted PIMC simulations which can be tested against our data. This might help to extend the range of validity of those simulations to higher density and lower temperature. Since this latter method does not have a sign problem it may allow to reach parameters that are not accessible otherwise. 
  \item 
  A major challenge of Metropolois-based QMC simulations that are highly efficient for thermodynamic and static properties is to extend them to dynamic quantities. This can, in principle, be done via analytical continuation from imaginary to real times (or frequencies). However, this is known to be an ill-posed problem. Recently, there has been significant progress by invoking stochastic reconstruction methods or genetic algorithms. For example, for Bose systems, accurate results for the spectral function and the dynamics structure factor could be obtained, e.g.~\cite{filinov_pra_12} and references therein which are encouraging also for applications to the uniform electron gas, in the near future.
  \item
  Finally, there is a large number of additional applications of the presented {\em ab initio} simulations. This includes the 2D warm dense UEG where thermodynamic results of similar accuracy should be straightforwardly accessible. Moreover, for the electron gas, at high density, $r_s \lesssim 0.1$, relativistic corrections should be taken into account. Among the presented simulations, CPIMC is perfectly suited to tackle this task and to provide {\em ab initio} data also for correlated matter at extreme densities.
\end{enumerate}

\section*{Acknowledgements}
This work was supported by the Deutsche Forschungsgemeinschaft via project BO1366-10 and via SFB TR-24 project A9 as well as grant shp00015 for CPU time at the Norddeutscher Verbund f\"ur Hoch- und H\"ochstleistungsrechnen (HLRN).
TS~acknowledges the support of the US DOE/NNSA under Contract No.~DE-AC52-06NA25396.
FDM is funded by an Imperial College PhD Scholarship.
FDM and WMCF used computing facilities provided by the High Performance Computing Service of Imperial College London, by the Swiss National Supercomputing Centre (CSCS) under project ID s523, and by ARCHER, the UK National Supercomputing Service, under EPSRC grant EP/K038141/1 and via a RAP award.
FDM and WMCF acknowledge the research environment provided by the Thomas Young Centre under Grant No.~TYC-101.

\end{document}